\documentclass[traditabstract,useAMS,usenatbib]{aa} 
\usepackage{natbib,graphicx}
\usepackage{times}
\usepackage{txfonts}

\def\ch2{$\chi^2$}

\def\Lo{L$_\odot$}
\def\Mo{M$_\odot$}

\def\arcsec {\hbox{$^{\prime\prime}$}}

\def \AL {$\alpha $}     
\def \HI {H{\sc \,i}}
\def \carb {\hbox{[C{\sc \,ii}]}}

\def\lapp{\ifmmode\stackrel{<}{_{\sim}}\else$\stackrel{<}{_{\sim}}$\fi}
\def\gapp{\ifmmode\stackrel{>}{_{\sim}}\else$\stackrel{>}{_{\sim}}$\fi}
\begin{document}
   \title{Ionised carbon and galaxy activity}

   \subtitle{}

   \author{S. J. Curran
          }

   \institute{School of Physics, University of New
  South Wales, Sydney NSW 2052, Australia\\
              \email{sjc@phys.unsw.edu.au}
         }
            
   \date{}

 
  \abstract{We investigate the possibility that the decrease in the
    relative luminosity of the 158~$\mu$m [C{\sc \,ii}] line with the
    far-infrared luminosity in extragalactic sources stems from a
    stronger contribution from the heated dust emission in the more
    distant sources. Due to the flux limited nature of these surveys,
    the luminosity of the detected objects increases with distance.
    However, the \carb\ luminosity does not climb as
    steeply as that of the far-infrared, giving the decline in the
    $L_{\rm [CII]}/L_{\rm FIR}$ ratio with $L_{\rm
      FIR}$. Investigating this further, we find that the
    \carb\ luminosity exhibits similar drops as measured against the
    carbon monoxide and radio continuum luminosities. The former may
    indicate that at higher luminosities a larger fraction of the
    carbon is locked up in the form of molecules and/or that the CO
    line radiation also contributes to the cooling, done mainly by the
    [C{\sc \,ii}] line at low luminosities. The latter hints at
    increased activity in these galaxies at greater distances, so we
    suggest that, in addition to an underlying heating of the dust by a
    stellar population, there is also heating of the embedded dusty
    torus by the ultra-violet emission from the active nucleus,
    resulting in an excess in the far-infrared emission from the more
    luminous objects. 

   \keywords{galaxies: abundances -- galaxies: ISM  -- galaxies: active -- Infrared: galaxies -- Radio continuum: galaxies-- quasars: emission lines 
               }}

   \maketitle
%
\section{Introduction}

\begin{figure}
 \centering
\includegraphics[angle=270,scale=0.75]{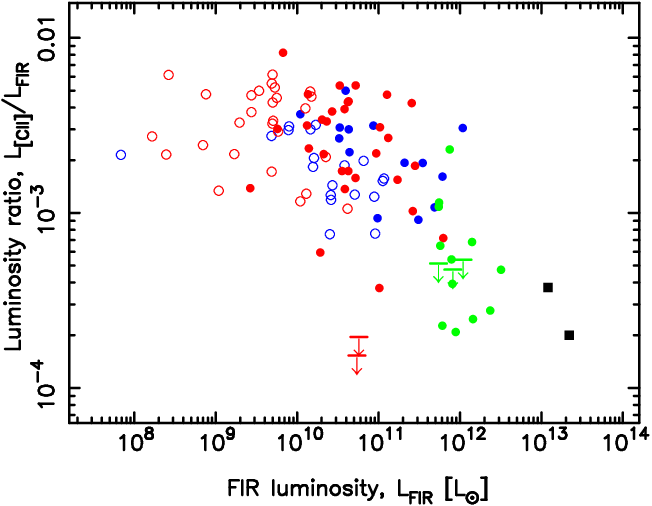}
\caption{The $L_{\rm [CII]}/L_{\rm FIR}$ ratio versus $L_{\rm FIR}$
  for the low redshift ($z\leq0.1296$) galaxies and the high redshift quasar
  searches (cf. figure 2 of \citealt{mcc+05}). Throughout this paper
  the symbols are colour coded according to the source reference --
  red \citep{mkh+01}, blue \citep{nocr01}, green \citep{lsf+03} and
  black (the two high redshift \carb\ detections of
  \citealt{mcc+05,iye+06}). The downwards arrows show
  the $3\sigma$ upper limits. As per \citet{lsf+03,mcc+05}, the hollow symbols
  indicate where the $\approx80\arcsec$ LWS aperture
  (\citealt{caa+96}) subtends $\lapp10$ kpc and the filled symbols
  where the aperture subtends $\gapp10$ kpc (at angular diameter
  distances of $\gapp26$ Mpc or $z\gapp0.006$ -- throughout this paper
  we use $H_{0}=71$~km~s$^{-1}$~Mpc$^{-1}$, $\Omega_{\rm matter}=0.27$
  and $\Omega_{\Lambda}=0.73,$ \citealt{svp+03}).}
\label{ratio-fir}
    \end{figure}

 The $^2{\rm P}_{3/2}\rightarrow^2{\rm P}_{1/2}$ fine-structure line of
    C$^+$, \carb, is believed to be a cooling pathway for the diffuse
gas in galaxies \citep{dm72}. This transition traces
    photo-dissociation regions (PDRs), where the ultra-violet
    radiation from young stars dominates the heating of the gas.  From
    this process, the \carb\ luminosity may reach up to 1\% the total
    luminosity of the galaxy \citep{cgtw85,sgg+91,wmb+91}, thus being
    the most powerful emission line in many galaxies. From a study of
60 normal galaxies with the Long Wavelength Spectrometer (LWS)
    on-board the Infrared Space Observatory (ISO, \citealt{caa+96}),
    \citet{mkh+01} find that for far-infrared luminosities of $L_{\rm
    FIR}\lapp10^{11}$ \Lo, the $L_{\rm [CII]}/L_{\rm FIR}$ ratio is
    roughly constant at $\log_{10}(L_{\rm [CII]}/L_{\rm
    FIR})\sim-2.5$, but drops rapidly above these values
    (Fig. \ref{ratio-fir}). This normalised decrease in
    the brighter galaxies is attributed to
    an increased grain charge, lowering
    the kinetic energy of the liberated photo-electrons 
    which deliver heat to the gas.

\citet{nocr01} extended this sample to include starburst galaxies and
active galactic nuclei (AGN), all of which are found to follow the
same trend, suggesting that far-infrared emission in the active
galaxies also arises primarily from star forming activity. Rather than
an increase in the charge carried by the dust grains, \citet{nocr01}
suggest that the $L_{\rm [CII]}/L_{\rm FIR}$ decrease with $L_{\rm
  FIR}$ is due to higher gas densities resulting in higher collision
rates, thus de-exciting the ionised carbon through a non-radiative
process.

\citet{lsf+98,lsf+03} confirm the decrease in $L_{\rm [CII]}/L_{\rm FIR}$
with $L_{\rm FIR}$ to higher luminosities ($L_{\rm FIR}\gapp10^{12}$
\Lo), by observing \carb\ in a sample of ultra-luminous infrared
galaxies (ULIRGs). This trend is in part attributed to much of the
far-infrared emission arising from dust-bounded photo-ionised gas which
does not contribute to the \carb\ emission.  Most recently, the
far-infrared luminosities have been taken up another notch by the two
high redshift detections of \carb, which again follow the same decline
in $L_{\rm [CII]}/L_{\rm FIR}$ \citep{mcc+05,iye+06}. Once more, this
indicates different excitation conditions than in local galaxies
\citep{iye+06} and possibly extremely high star formation rates
($\sim3000$ \Mo\ yr$^{-1}$, \citealt{mcc+05}) in the early
Universe. From all of these studies (summarised in
Fig. \ref{ratio-fir}), there is no doubt that the relative strength of
the \carb\ line drops with far-infrared luminosity. However, such flux
limited surveys are subject to a selection effect, where only the
brighter sources are detected at large distances.  It is therefore
possible that the \carb\ deficit is caused by changing demographics of
the galaxies at larger distances, a possibility we investigate in this
paper.

\section{Possible selection effects}

\subsection{Relative \carb\ luminosities}
\label{rcl}

\begin{figure}
\centering
\includegraphics[angle=270,scale=0.75]{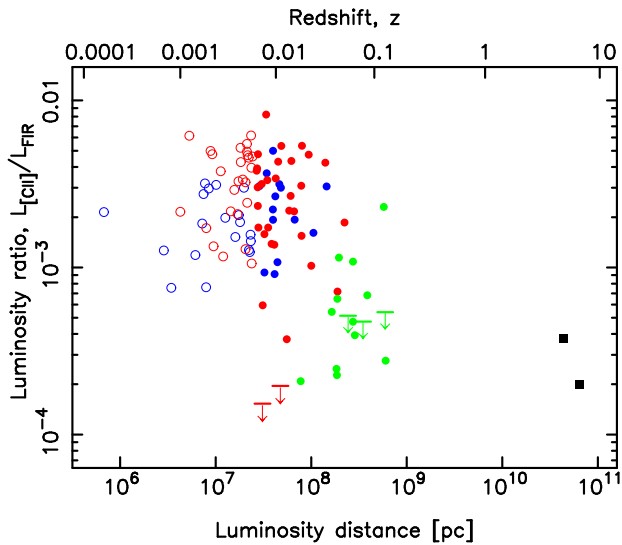}
\caption{The $L_{\rm [CII]}/L_{\rm FIR}$ ratio versus the luminosity
  distance. The symbols are as per Fig. \ref{ratio-fir}.}
\label{ratio-lbt}
\end{figure}

In Fig. \ref{ratio-lbt} we replot the \carb--FIR luminosity ratio
against the luminosity distance (cf. Fig. \ref{ratio-fir}),
from which we find a very similar trend. Using Kendall's $\tau$ rank
correlation coefficient, which is a non-parametric test of the degree
of correspondence between two parameters, we find a 0.009 probability
that there is no correlation between the $L_{\rm [CII]}/L_{\rm FIR}$
ratio and distance (Table~\ref{stats}\footnote{In Table~\ref{stats}
  the upper limits are incorporated according to the survival analysis
  of \citet{ifn86}, via the {\sc asurv} package. This gives
  Kendall's $\tau$ two-sided probability that there is no
  correlation between the two ranks, $P(\tau)$, and the subsequent
  significance of the correlation, $S(\tau)$.
  Note that where 
  $n=108$, we have had to exclude the two blue-shifted galaxies
  (NGC\,1569; \citealt{mkh+01} and Maffei\,2; \citealt{nocr01}), since
  the tested parameters must be converted to $\log$ values to run {\sc
    asurv}. With $L_{\rm [CII]}/L_{\rm FIR}\gapp 10^{-3}$, these
  occupy the same regions as the other $z<0.006$ points and do not
  weaken the correlations.}). This decreases to $P(\tau)=0.002$ for
the LWS ``point'' sources (i.e. those at $z>0.006$), where the
\carb\ emission should be fully sampled.
\begin{table}
\caption{Sample statistics for various redshift ranges -- (1) the
  whole sample, (2) the sources for which the LWS aperture subtends
  $\gapp10$ kpc, (3) also excluding the two high-redshift quasars
  \citep{mcc+05,iye+06} and finally, (4) the sources for which the
  aperture subtends $\lapp10$~kpc.}
\label{stats}
\centering
\begin{tabular}{l c c c} 
\hline\hline
\noalign{\smallskip}
Redshift range & $n$ & P($\tau$) & S($\tau$)\\
\noalign{\smallskip}
\hline
\noalign{\smallskip}
\multicolumn{4}{c}{$L_{\rm [CII]}/L_{\rm FIR}$--Luminosity distance (Fig. \ref{ratio-lbt})}\\
\noalign{\smallskip}
\hline
\noalign{\smallskip}
Whole & 108 &0.0090 & $2.61\sigma$ \\
$z>0.006$ & 63 &0.0022 &  $3.06\sigma$ \\
$0.006\leq z\leq0.1296$ & 61 & 0.0114  & $2.53\sigma$ \\
$0<z\leq0.006$ & 45 & 0.1589 & $1.41\sigma$ \\
\noalign{\smallskip}
%
\hline
\noalign{\smallskip}
\multicolumn{4}{c}{$L_{\rm [CII]}-L_{\rm FIR}$ (Fig. \ref{cii-fir})}\\
\noalign{\smallskip}
\hline
\noalign{\smallskip}
Whole & 110 & $5.9\times10^{-30}$ & $11.4\sigma$ \\
$z>0.006$ & 63 &$1.1\times10^{-14}$ &  $7.73\sigma$ \\
$0.006<z\leq0.1296$ & 61 & $1.8\times10^{-13}$ & $7.36\sigma$ \\
$0<z\leq0.006$ & 45 & $5.2\times10^{-13}$ & $7.22\sigma$ \\
\noalign{\smallskip}
\hline
\noalign{\smallskip}
\multicolumn{4}{c}{$L_{\rm [CII]}$--Luminosity distance (Fig. \ref{cii-fir-lbt})}\\
\noalign{\smallskip}
\hline
\noalign{\smallskip}
Whole & 108 & $1.1\times10^{-19}$ & $9.08\sigma$ \\
$z>0.006$ & 63 &$4.1\times10^{-10}$ &  $6.25\sigma$ \\
$0.006<z\leq0.1296$ & 61 & $6.6\times10^{-9}$ & $5.80\sigma$ \\
$0<z\leq0.006$ & 45 & 0.0060 & $2.75\sigma$ \\
\noalign{\smallskip}
\hline
\noalign{\smallskip}
\multicolumn{4}{c}{$L_{\rm FIR}$--Luminosity distance (Fig. \ref{cii-fir-lbt})}\\
\noalign{\smallskip}
\hline
Whole & 108 & $2.3\times10^{-19}$ & $9.00\sigma$ \\
$z>0.006$ & 63 &$8.1\times10^{-13}$ & $7.16\sigma$ \\
$0.006<z\leq0.1296$ & 61 & $1.6\times10^{-11}$ & $6.74\sigma$ \\
$0<z\leq0.006$ & 45 & 0.1152 & $1.58\sigma$ \\
\noalign{\smallskip}
\hline
\end{tabular}
\end{table}
If the $L_{\rm [CII]}/L_{\rm FIR}$ ratio and the distance are
unrelated, a 0.2\% probability would be located at $3.06\sigma$ on the
tails of a normalised Gaussian (mean$\,=0$, $\sigma=1$), suggesting
that there is a correlation.

\begin{figure}
\centering
\includegraphics[angle=270,scale=0.75]{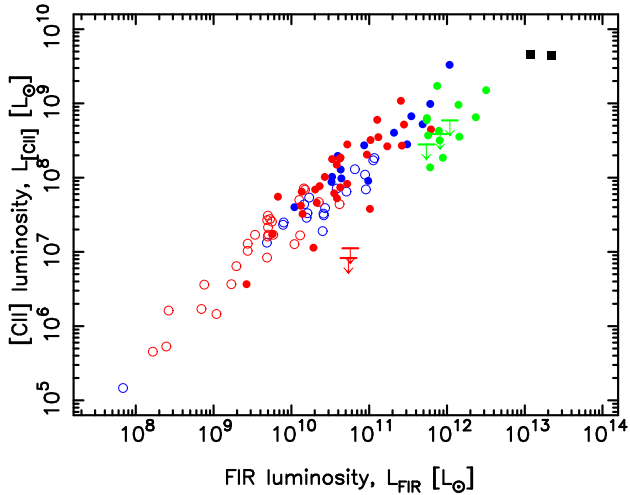}
\caption{The \carb\ luminosity versus the FIR luminosity.}
\label{cii-fir}
\end{figure}

Clearly, there exists the possibility that the correlation is
dominated by the inclusion of the two high-redshift quasars ($z =
6.42$, \citealt{mcc+05} and $z=4.69$, \citealt{iye+06},
cf. $z\leq0.1296$ for the rest of the sample). Excluding these from
the statistics, however, we see that a correlation remains at a
$2.53\sigma$ significance.  Finally, note that at $z = 0.0072$ and
$0.011$, the two non-detections of \citet{mkh+01}, which are the only
two significant outliers (as they are for the $L_{\rm [CII]}/L_{\rm
  FIR}$ versus $L_{\rm FIR}$ correlation, Fig. \ref{ratio-fir}), 
lie close to the $z=0.006$ ``beam-filling'' cut-off. At these
respective redshifts, the $<80\arcsec$ LWS beam \citep{nocr01}
subtends $<11$ and $<17$ kpc and so these may not represent true
$L_{\rm [CII]}$ upper limits (although cf. Fig. \ref{cii-fir}). The
exclusion of these sources raises the significances to $2.89\sigma$
(whole sample) and $3.74\sigma$ ($z>0.006$). We include these in
Table \ref{stats}, however, as per the other authors, we use a cut-off
of 10~kpc for the diameter of the LWS beam. Finally, there is also the
possibility of an undetected population of objects with large $L_{\rm
  [CII]}/L_{\rm FIR}$ ratios at large distances (i.e. located to the
top right of Fig. \ref{ratio-lbt}). If there is a population of faint
sources at large distances, their effect on Fig. \ref{ratio-fir} would
have to be considered, although these would suggest that the decline
in $L_{\rm [CII]}/L_{\rm FIR}$ with luminosity distance is a selection
effect, introduced by the flux limited nature of the surveys.  Caution
must therefore be advised in interpreting Fig. \ref{ratio-fir} and the
following figures, as it is possible that the locus of points may
actually represent the upper luminosity edge of a wedge of undetected
lower luminosity sources.

\begin{figure}
\centering
\includegraphics[angle=270,scale=0.75]{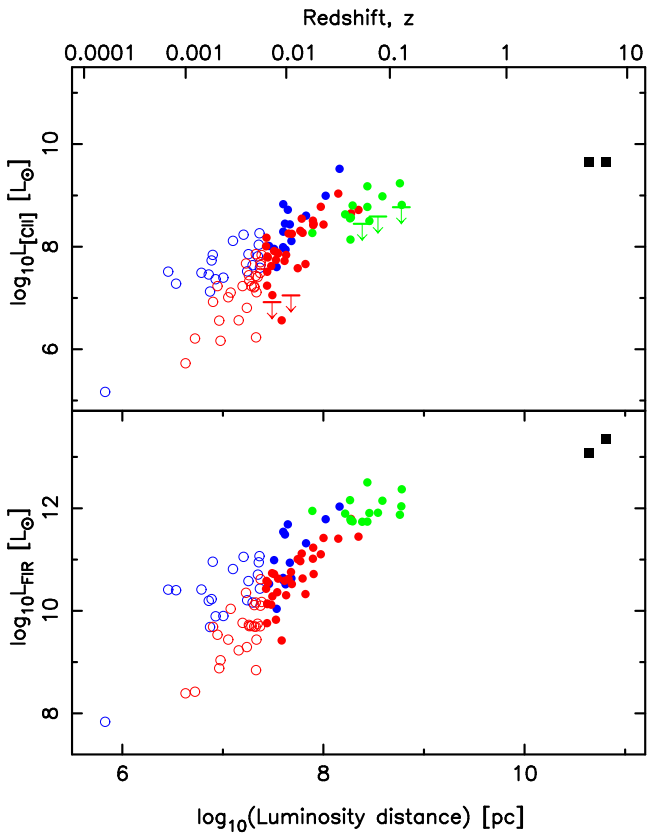}
\caption{The \carb\ and FIR luminosities versus the luminosity distance.
  The least-squares fit to $\log_{10} L_{\rm
    [CII]}$--$\log_{10}\,$(luminosity distance) has a gradient of
  $0.89\pm0.07$ (incorporating the upper limits) and the fit to
  $\log_{10} L_{\rm FIR}$--$\log_{10}\,$(luminosity distance) has a
  gradient of $1.16\pm0.08$, over the whole redshift range. The
  panels are plotted to cover the same luminosity range (in decades),
 from which we see a clear deficit in $L_{\rm
    [CII]}$ in comparison to $L_{\rm FIR}$ as the distances increase.}
\label{cii-fir-lbt}
\end{figure}

Fig.~\ref{cii-fir} shows the \carb\ versus the FIR luminosity,
which appear to be closely
correlated over the whole range of luminosities
(Table~\ref{stats}). Note, however, that with a gradient of less
than unity ($0.77\pm0.03$, incorporating the upper limits), the
least-squares fit to these data demonstrates that $L_{\rm [CII]}$ does
not match the climb in $L_{\rm FIR}$, confirming a
depletion in the \carb\ luminosity in relation to the FIR
emission. This is further illustrated in Fig.~\ref{cii-fir-lbt},
where the increase in both $L_{\rm [CII]}$ and $L_{\rm FIR}$ with
distance is due to the increase in the lower limit of the luminosities
which can be detected. The figure confirms that $L_{\rm [CII]}$ does
not climb quite so rapidly as $L_{\rm FIR}$, with an apparent slowing
in the increase of $L_{\rm [CII]}$ at $\gapp10^8$ pc, where the ULIRGs
dominate. This indicates that the $L_{\rm [CII]}/L_{\rm
  FIR}$--luminosity distance trend (Fig. \ref{ratio-lbt}) is the result of lower
relative $L_{\rm [CII]}$ contribution at larger distances and that the
anti-correlation is not purely due to the two high redshift points.

\subsection{Possible causes}

\subsubsection{Increasing AGN activity}
\label{agn}

Even with the exclusion of the two high redshift quasars the redshift
range spanned by the sample is considerable, with $z\leq0.1296$
corresponding to a luminosity distance of $\leq600$ Mpc, or looking
back 12\% into the history of the Universe. At these moderate
redshifts, we may expect a change in the demographics of the galaxies
from those in the local Universe, as is seen from the presence of the
ULIRGs (above) or a larger population of AGN (see below). In fact,
among a list of scenarios possibly responsible for the decline in
$L_{\rm [CII]}/L_{\rm FIR}$, \citet{mkh+01} have suggested that
increased AGN activity, over and above that of the starburst, could be
the cause of the changing $L_{\rm [CII]}/L_{\rm FIR}$ ratios.

\begin{figure}
\centering 
\includegraphics[angle=270,scale=0.75]{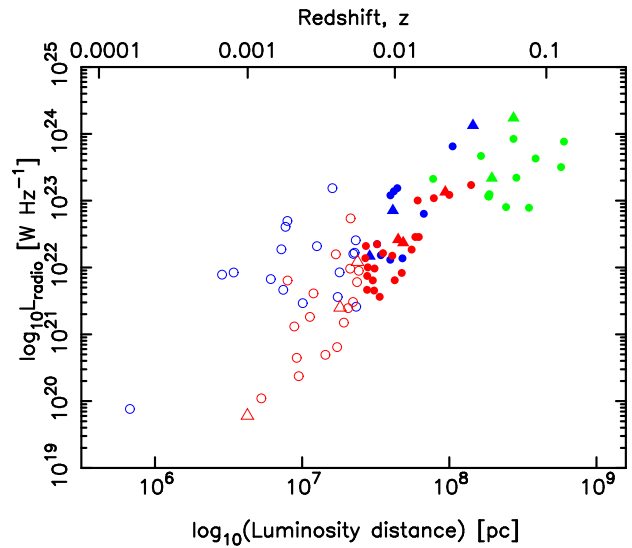}
\caption{The 1.4 GHz continuum luminosity versus the luminosity
  distance. We show this up to $z\sim0.13$, since there are no radio
  flux measurements available for the two high redshift sources. The
  triangles designate fluxes interpolated from neighbouring radio
  frequencies. Note that the top right corner is within the realm of
  radio galaxies and quasars where luminosities of $L_{\rm
    radio}\gapp10^{24}$ W Hz$^{-1}$ are found at $z\gapp0.1$ (see
  figure 4 of \citealt{cww+08}.)}
\label{radio-lbt}
\end{figure}

AGN activity may be traced by radio loudness, with radio surveys at
1.4~GHz finding a bimodal distribution in the brightness of
extragalactic radio sources: The vast majority of the radio-loud
(over 95\% with $S_{\rm radio}\gapp50$ mJy) being radio galaxies and
quasars, whereas the radio-quiet tend to be star-forming galaxies
(which dominate at flux densities of $S_{\rm radio}\lapp1$ mJy,
\citealt{con84,wmo+85}). We have therefore trawled the NASA/IPAC
Extragalactic Database (NED) for the 1.4 GHz flux densities of the
galaxies searched for in \carb\ and converted these to radio
luminosities, which we show in Fig.~\ref{radio-lbt}.
As seen from this (and Table \ref{stats2}), there is a very strong
correlation between radio luminosity and distance. Again, this
is not surprising due to the flux limited nature of these surveys, but it
does show that even over the low redshift sample ($z\leq0.1296$), there is
a strong selection effect. 

\begin{figure}
\centering \includegraphics[angle=270,scale=0.75]{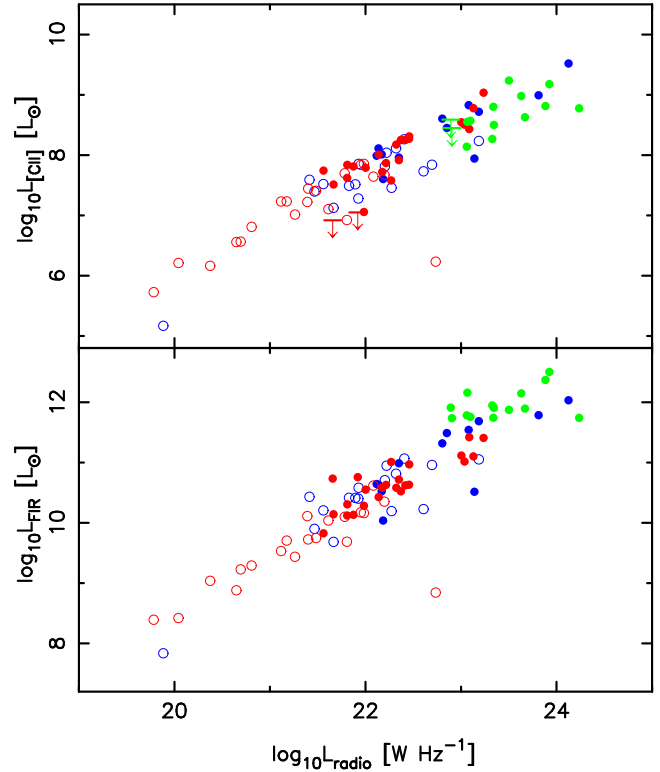}
\caption{The \carb\ and FIR luminosities versus the 1.4 GHz continuum
  luminosity.  The least-squares fit to $\log_{10} L_{\rm [CII]} -
  \log_{10} L_{\rm radio}$ has a gradient of $0.77\pm0.04$
  (incorporating the upper limits) and the fit to $\log_{10} L_{\rm
    FIR} - \log_{10} L_{\rm radio}$ has a gradient of $0.92\pm0.05$,
  over the whole radio luminosity range. As per Fig. \ref{cii-fir-lbt},
  the ordinate in both panels are plotted to cover the same luminosity
  range (in decades).}
\label{cii-fir-radio}
\end{figure}

Along with the distinct differences in radio fluxes, there is a
difference in the redshift distributions, with the AGN exhibiting
the higher values \citep{con84}. This is confirmed by the 2dF and 6dF
Galaxy Redshift Surveys (\citealt{smjc99,ms07}), where star-forming
galaxies have a median redshift of $z\approx0.05$, in contrast to $z\approx0.1$
for the AGN, which also exhibit a longer high redshift tail (up to the
$z=0.3$ limit of the surveys)\footnote{Note also, from X-ray
  photometry \citet{zmm+04} find that the galaxy population drops at
  $z\gapp1$, in comparison to $z\gapp2$ for type-2 AGN and no significant redshift
  dependence for type-1 AGN.  From rest-frame UV photometry,
  \citet{cww+08} also suspect that all optical+radio bright sources at
  $z\gapp3$ are type-1 AGN.}. Over the range of this sample
($z\leq0.1296$), the vast majority of star forming galaxies in the 2dF
sample have radio luminosities of $L_{\rm
  radio}\lapp10^{23}$~W~Hz$^{-1}$, with most AGN kicking in at
$z\gapp0.1$ with $L_{\rm radio}\gapp10^{23}$~W~Hz$^{-1}$
\citep{smjc99}\footnote{The range of radio luminosities found may
  suggest a range of $\approx0.07$ to $\approx0.3\times10^{9}$\Mo\ for
  the masses of the central black holes powering the galaxies
  \citep{mm06}, over luminosity distances of $\sim10^{7}$ to $\sim10^{9}$
 pc (Fig.~\ref{radio-lbt}).}. Note that for the latter this value
is close to the radio flux limit, and so a more radio-faint AGN
population at high redshift cannot be ruled out. However, 87 of the
108 sources have published (and detected) radio fluxes, where radio
loudness is not necessarily a prerequisite for far-infrared selected
surveys\footnote{\citet{mkh+01} and \citet{nocr01} select near-by
  normal star-forming galaxies and \citet{lsf+03} select ULIRGs.},
although ionised gas will emit a radio continuum (see \citealt{nocr01}).  As
well as the possibility of a faint AGN population, there are expected
to be higher redshift star forming galaxies, but by $z\approx0.2$,
these are already below the $\sim$mJy detection threshold. As stated
above, most of the \carb\ sample is detected in the radio, due in part
to the generally low redshifts, and in Fig. \ref{cii-fir-radio} we
show the \carb\ and FIR luminosities against that of the radio.

From this, we see that \carb\ luminosity does not climb as steeply as
that of the FIR with radio luminosity, which is clearly apparent even
with the exclusion of the two high redshift sources
(cf. Fig.~\ref{cii-fir-lbt}). Rather than heating by stars, the
correlation of $L_{\rm FIR}$ with $L_{\rm radio}$ may suggest a
significant AGN contribution, where much of the FIR emission may arise
from dust heated by ultra-violet emission from the central accretion
disk. This is also suspected to be the case in a sample of low
redshift Seyfert galaxies, where the FIR emission does not wholly
trace the dense star-forming molecular cores
\citep{cjhb99,capb01b}. In extreme cases ($L_{\rm
  UV}\gapp10^{23}$~W~Hz$^{-1}$), high ultra-violet fluxes may be
reponsible for ionising much of the neutral gas \citep{cww+08}, making
star formation, ironically enough, less likely in the most UV bright
sources.  Note finally that, although there are no radio fluxes
available for the two high redshift \carb\ detections, the presence of
a powerful X-ray flux from the quasar is invoked by \citet{mcc+05} to
account for the large populations in the high CO rotational levels at
$z = 6.42$, which cannot be obtained from a PDR model of a typical
star forming region alone.

\subsubsection{Decreasing metallicities}
\label{metal}

As mentioned above, the galaxies searched probe the past 12\% of the
Universe, with the two high redshift detections providing end-points
at look-back times of 12.4 and 12.8 Gyr, i.e.  within the first 12\%
of the Universe's lifetime. An evolutionary effect, which may give the
observed decrease in the \carb\ line emission fraction, could be the
cosmological evolution of heavy element abundances. A correlation
between metallicity and look-back time has already been observed in
damped Lyman-\AL\ absorption systems over the first 6 Gyr history of
the Universe \citep{pgw+03,cwmc03}, and an increase in the carbon
abundance with cosmic time could explain the lower relative
\carb\ contribution with increasing redshift (at least over the full
$0 < z < 6.42$). A local low metallicity laboratory is the Magellanic
clouds, and in the 30 Doradus region of the LMC, \citet{sgg+91} find
\carb/CO intensity ratios $\approx30$ times larger than Galactic
values, later confirmed to be a factor of $\approx20$ over the 
main part of the LMC ($\approx6\times10$ kpc$^2$, \citealt{mnd+94}). This is
interpreted as the lower metallicities yielding lower dust
abundances\footnote{Such a correlation has been noted at high redshift
  by \citet{cwmc03}.} and thus providing less shielding from
ultra-violet photons, dissociating and ionising the CO into \carb. In
this model the high \carb/CO ratios are therefore indicative of low
metallicities.

\begin{figure}
\centering
\includegraphics[angle=270,scale=0.75]{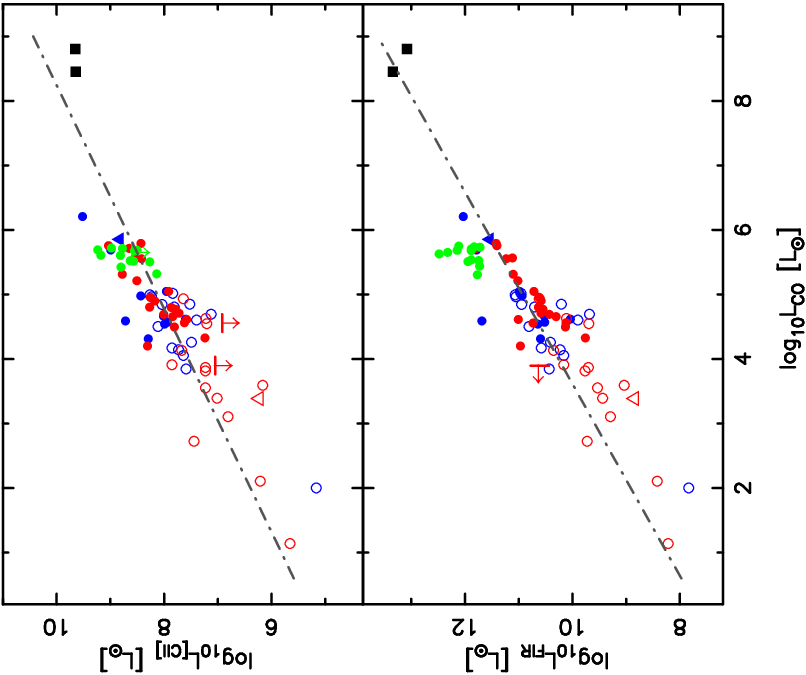}
\caption{The \carb\ and FIR luminosities versus the CO luminosities,
  which have been compiled from
  \citet{sm85,ssy+86,sss88a,sss91,ss88,hbw+89,wh89,ssp89,ecr+90,lkrp91,tttv91,cs92,sag93,abbj95,yxt+95,gbjm96,ebh+96,opg+96,sdrb97,gs99,capb01a,blg02,dwm02,bcn+03,yskd03,spa+04,lbsb05,akc07}. The
  shapes designate the CO rotational transition; circles -- $J =
  1\rightarrow0$, triangles -- $J = 2\rightarrow1$ and the squares the
  higher CO transitions for the two high redshift
  \carb\ detections. The lines show the least-squares fits to all of
  the points, in which $\log_{10} L_{\rm [CII]}-\log_{10} L_{\rm CO}$
  has a gradient of $0.54\pm0.05$ and $\log_{10} L_{\rm FIR}-\log_{10}
  L_{\rm CO}$ a gradient of $0.67\pm0.05$.}
\label{cii-fir-co}
\end{figure}
Plotting the \carb\ and FIR luminosities against that of CO
(Fig.~\ref{cii-fir-co}), we see that both are strongly correlated with
this tracer of molecular gas abundance (Table \ref{stats2}) and, again,
the \carb\ luminosity does not climb as rapidly as $L_{\rm FIR}$ with increasing
$L_{\rm CO}$. Over this wide range of luminosities (and redshifts),
there is no large change in the $L_{\rm [CII]}-L_{\rm CO}$ ratio
apparent: From a recent survey of the LMC, \citet{bji+00} detect
more extended CO emission, undetected by \citet{mnd+94}, the
contribution of which brings the $L_{\rm [CII]}/L_{\rm CO}$ ratio
close to Galactic values. Further afield, \citet{mpg+97} find widely
varying $L_{\rm [CII]}/L_{\rm CO}$ ratios in the low-metallicity dwarf
galaxy IC\,10 and \citet{sm97} find in two spiral galaxies $L_{\rm
[CII]}/L_{\rm CO}$ ratios which are an order of magnitude higher than
Galactic disk values and more typical of the values found in irregular
(low metallicity) galaxies. This calls into question the effectiveness
of this ratio as a tracer of heavy element abundance (although see
\citealt{bji99,roj+06}).
\begin{table}
\caption{As per Table \ref{stats}.}
\label{stats2}
\centering
\begin{tabular}{l c c c} 
\hline\hline
\noalign{\smallskip}
Redshift range & $n$ & P($\tau$) & S($\tau$)\\
\noalign{\smallskip}
\hline
\noalign{\smallskip}
\multicolumn{4}{c}{$L_{\rm radio}$--Luminosity distance (Fig. \ref{radio-lbt})}\\
\noalign{\smallskip}
\hline
\noalign{\smallskip}
Whole & 87 & $9.7\times10^{-16}$ & $8.03\sigma$ \\
$0.006\leq z\leq0.1296$ & 50 &$3.6\times10^{-10}$ & $6.27\sigma$ \\
$0<z\leq0.006$ &37 &0.0891 &  $1.70\sigma$ \\
\noalign{\smallskip}
\hline
\noalign{\smallskip}
\multicolumn{4}{c}{$L_{\rm [CII]} - L_{\rm radio}$ (Fig. \ref{cii-fir-radio})}\\
\noalign{\smallskip}
\hline
\noalign{\smallskip}
Whole & 87 & $1.9\times10^{-23}$ & $9.98\sigma$ \\
$0.006<z\leq0.1296$ & 50 & $7.5\times10^{-13}$ & $7.17\sigma$ \\
$0<z\leq0.006$ & 37 &$2.9\times10^{-8}$  & $5.55\sigma$ \\
\noalign{\smallskip}
\hline
\noalign{\smallskip}
\multicolumn{4}{c}{$L_{\rm FIR} - L_{\rm radio}$ (Fig. \ref{cii-fir-radio})}\\
\noalign{\smallskip}
\hline
\noalign{\smallskip}
Whole & 87 & $5.5\times10^{-24}$ & $10.1\sigma$ \\
$0.006<z\leq0.1296$ & 50 & $1.3\times10^{-11}$ & $6.77\sigma$ \\
$0<z\leq0.006$ & 37 &$1.1\times10^{-8}$  & $5.72\sigma$ \\
\noalign{\smallskip}
\hline
\noalign{\smallskip}
\multicolumn{4}{c}{$L_{\rm [CII]}-L_{\rm CO}$ (Fig. \ref{cii-fir-co})}\\
\noalign{\smallskip}
\hline
\noalign{\smallskip}
Whole &80 &$9.4\times10^{-19}$ & $8.84\sigma$\\
$z>0.006$ & 49& $1.6\times10^{-10}$& $6.40\sigma$\\
$0.006<z\leq0.1296$ &47 & $3.0\times10^{-9}$& $5.93\sigma$\\
$0<z\leq0.006$ & 30& 0.000055& $4.03\sigma$\\
\noalign{\smallskip}
\hline
\noalign{\smallskip}
\multicolumn{4}{c}{$L_{\rm FIR}-L_{\rm CO}$ (Fig. \ref{cii-fir-co})}\\
\noalign{\smallskip}
\hline
\noalign{\smallskip}
Whole  &80 &$1.4\times10^{-20}$ & $9.03\sigma$\\
$z>0.006$ & 49 & $2.8\times10^{-10}$& $6.31\sigma$\\
$0.006<z\leq0.1296$& 47& $4.6\times10^{-9}$& $5.86\sigma$\\
$0<z\leq0.006$ & 30 & $1.3\times10^{-6}$& $4.84\sigma$\\
\noalign{\smallskip}
\hline
\end{tabular}
\end{table}

\subsubsection{Star formation rates}

A correlation between the \carb\ and CO intensities has previously
been noted by \citet{sgg+91}, leading to the hypothesis that the
ionised carbon and carbon monoxide are spatially
coincident. \citet{abbj95} suggested that the CO must be reasonably
excited (so that $I_{{\rm CO}~2\rightarrow1}/I_{{\rm
    CO}~1\rightarrow0}\gapp0.8$) if associated with a PDR and
demonstrated that such CO intensity ratios were satisfied for
\carb/CO$\,\gapp4000$ in a sample of 19 normal and starburst galaxies.
It therefore appears that large $L_{\rm [CII]}/L_{\rm CO}$ ratios are
indicative of enhanced star formation (\citealt{sgg+91}, see also
\citealt{bji+00}) and, from a sample of 21 late-type galaxies,
\citet{pltv99} find that $L_{\rm [CII]}/L_{\rm CO}$ is proportional to
the star-formation rate in non-starburst galaxies. Furthermore,
\citet{sgg+91} suggest that starburst galaxies and star forming
regions have ratios of \carb/CO$\,\approx6000$, three times higher
than for quiescent Galactic regions and non-starburst galaxies. In
Fig.~\ref{cii-fir-co}, apart from the $z\lapp0.006$ scatter and
possibly the ULIRGs, we see no major deviations from the $L_{\rm
  [CII]}/L_{\rm CO}$ trend, although the log plot will be quite
insensitive to a factor of three.  It is clear, however, that the
ULIRGs have systematically higher $L_{\rm FIR}/L_{\rm CO}$ ratios than
the rest of the sample, perhaps indicating a contribution to the FIR
luminosity from an AGN in addition to that from the heating by
stars. Note that \citet{capb01b} and \citet{gs04} find $L_{\rm
  FIR}\propto L_{\rm HCN}$ over three orders of magnitude of
luminosity, including the ULIRGs. This suggests that, while the CO
traces all of the molecular gas, the HCN, which traces the dense gas,
is closely associated with the FIR emission, be this due to star
formation and/or AGN activity \citep{capb01b}.

\subsubsection{Gas cooling by [O{\sc \,i}]}
\label{ogc}

In increased far ultra-violet fields the efficiency of the
photo-ejection of electrons from dust grains is reduced, whereas the
FIR dust emission continues to increase linearly. We may therefore
expect the observed relative decline in the \carb\ luminosities with
those of the FIR. However, since the increased luminosities may also
be tracing different galaxy types, we may also expect different
contributions from other line coolants.
\begin{figure}
\centering \includegraphics[angle=270,scale=0.75]{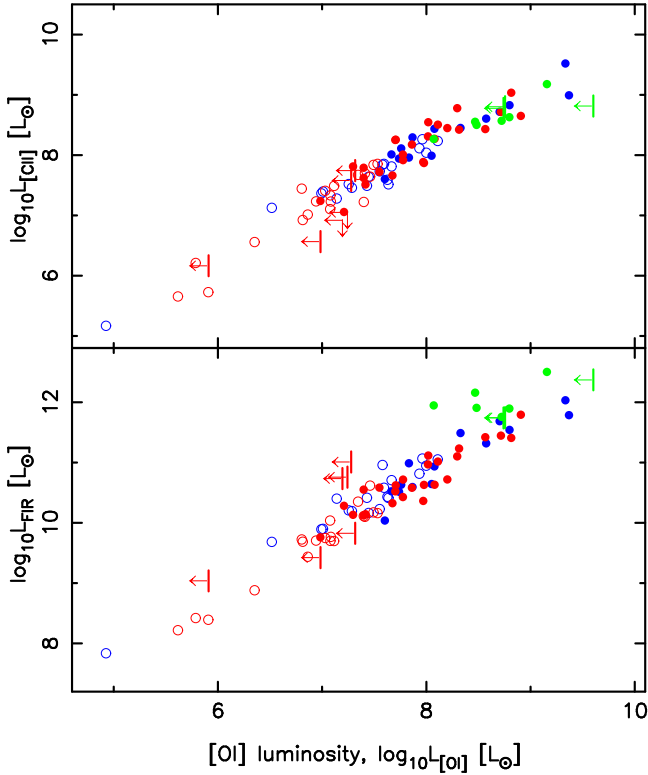}
\caption{The \carb\ and FIR luminosities versus the [O{\sc \,i}] 
  luminosity.  The least-squares fit to $\log_{10} L_{\rm [CII]} -
  \log_{10} L_{\rm [OI]}$ has a gradient of $0.87\pm0.05$
  (incorporating the upper limits) and the fit to $\log_{10} L_{\rm
    FIR} - \log_{10} L_{\rm [OI]}$ has a gradient of $1.05\pm0.08$,
  for $z > 0.006$. As per Fig. \ref{cii-fir-lbt},
  the ordinate in both panels are plotted to cover the same luminosity
  range (in decades).}
\label{cii-fir-OI}
\end{figure}
The other major coolant in galaxies is that of the [O{\sc \,i}] line,
which we show against the \carb\ and FIR luminosities in
Fig. \ref{cii-fir-OI}, where again we see a relative decline in $L_{\rm
  [CII]}$. A constant $L_{\rm [OI]}/L_{\rm FIR}$ ratio with $L_{\rm
  FIR}$ was previously noted by \citet{mkh+01,nocr01}, which is
interpreted as in increase in dust, and therefore gas, temperatures
due to a higher incident far ultra-violet flux \citep{kwhl99}.  Again,
we see that the FIR luminosities cause the ULIRGs to stray from the
lower luminosity trend, although the overall \carb--[O{\sc \,i}]
correlation holds tightly for these objects. As for the radio
continuum (Fig.~\ref{cii-fir-radio}) and molecular line
(Fig.~\ref{cii-fir-co}) emission, this indicates that there is an
excess of FIR emission at higher luminosities.

\begin{figure}
\centering
\includegraphics[angle=270,scale=0.75]{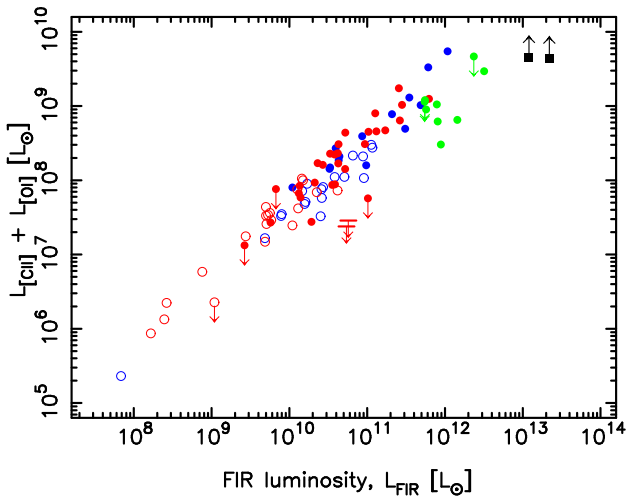}
\caption{The \carb\ + [O{\sc \,i}] luminosity versus the FIR luminosity. }
\label{cii+oi-fir}
\end{figure}

Lastly, in Fig. \ref{cii+oi-fir} we show the total coolant line luminosity
(cf. \citealt{mkh+01}) versus that of the far-infrared. 
The least-squares fit over the whole redshift range has a gradient of
$0.87\pm0.03$, which is higher than for the $L_{\rm [CII]}-L_{\rm
  FIR}$ correlation ($0.77\pm0.03$, Fig.~\ref{cii-fir}), although lower than
for $L_{\rm [OI]}-L_{\rm FIR}$ ($0.95\pm0.08$, Fig.~\ref{cii-fir-OI}).
This indicates that the [O{\sc \,i}] increases its contibution to the 
cooling of the gas with increasing luminosity and that the stifled \carb\ luminosity
has the effect of damping the total coolant line luminosity increase in relation
to the FIR. This could be partly responsible for the ULIRGs which are, once again, 
offset from the overall trend, although, as seen in Fig. \ref{cii-fir-OI},
there is also an FIR excess in relation to the [O{\sc \,i}]. Detections of
the [O{\sc \,i}]  emission in the two high redshift objects could verify
this trend at the highest luminosities.
\begin{table}[h]
\caption{As per Table \ref{stats}.}
\label{stats3}
\centering
\begin{tabular}{l c c c} 
\hline\hline
\noalign{\smallskip}
Redshift range & $n$ & P($\tau$) & S($\tau$)\\
\noalign{\smallskip}
\hline
\noalign{\smallskip}
\multicolumn{4}{c}{$L_{\rm [CII]} - L_{\rm [OI]}$ (Fig. \ref{cii-fir-OI})}\\
\noalign{\smallskip}
\hline
\noalign{\smallskip}
Whole & 94 & $8.3\times10^{-31}$ & $11.54\sigma$ \\
$0.006<z\leq0.1296$ & 54 & $5.1\times10^{-16}$ & $8.11\sigma$ \\
$0<z\leq0.006$ & 40 &$1.2\times10^{-12}$  & $7.11\sigma$ \\
\noalign{\smallskip}
\hline
\noalign{\smallskip}
\multicolumn{4}{c}{$L_{\rm FIR} - L_{\rm [OI]}$ (Fig. \ref{cii-fir-OI})}\\
\noalign{\smallskip}
\hline
\noalign{\smallskip}
Whole & 94 & $7.8\times10^{-26}$ & $10.51\sigma$ \\
$0.006<z\leq0.1296$ & 54 & $8.5\times10^{-12}$ & $6.83\sigma$ \\
$0<z\leq0.006$ & 40 &$4.5\times10^{-13}$  & $7.24\sigma$ \\
\noalign{\smallskip}
\hline
\noalign{\smallskip}
\multicolumn{4}{c}{$(L_{\rm [CII]} + L_{\rm [OI]}) - L_{\rm FIR}$ (Fig. \ref{cii+oi-fir})}\\
\noalign{\smallskip}
\hline
\noalign{\smallskip}
Whole & 96 & $5.1\times10^{-26}$ & $10.55\sigma$ \\
$z>0.006$ & 56 & $1.1\times10^{-12}$&$7.12\sigma$ \\ 
$0.006<z\leq0.1296$ & 54 & $1.8\times10^{-11}$ & $6.72\sigma$ \\
$0<z\leq0.006$ & 40 &$2.8\times10^{-12}$  & $6.99\sigma$ \\
\noalign{\smallskip}
\hline
\end{tabular}
\end{table}

\subsection{Recap of the correlations}

Here we replot the correlations in the same manner as Fig.~\ref{ratio-fir},
where we see similar trends as for $L_{\rm [CII]}/L_{\rm FIR} - L_{\rm FIR}$: 
\begin{figure}
 \centering
\includegraphics[angle=270,scale=0.75]{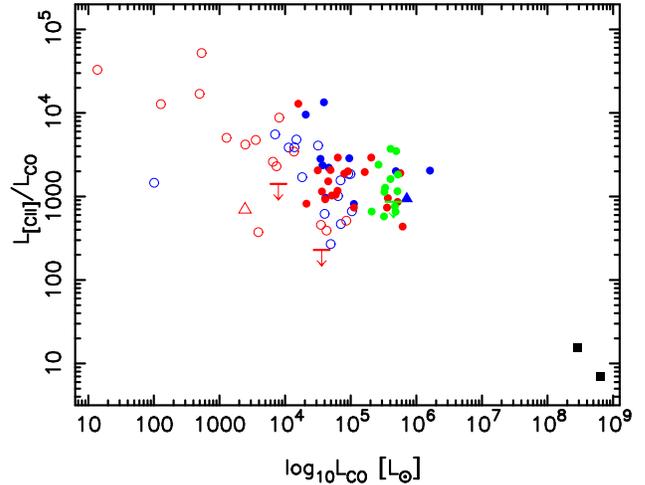}
\caption{As Fig. \ref{ratio-fir}, but normalising by the CO luminosity.}
\label{cii-over-co-co}
    \end{figure}
With the inclusion of the two high redshift points, the $L_{\rm
  [CII]}/L_{\rm CO} - L_{\rm CO}$ relation (Fig.~\ref{cii-over-co-co})
exhibits the steepest drop (four decades, cf. the not-quite two of the
FIR, Fig.~\ref{ratio-fir}). Since a significant contribution to this
decline is due to the two end points, this may reafirm our
belief that $L_{\rm [CII]}/L_{\rm CO}$ is a poor tracer of
metallicity, since we would expect the highest ratios (lowest
metallicities) for the high redshift sources (Sect. \ref{metal}). What
Fig.~\ref{cii-over-co-co} suggests, naturally enough, is that the
abundance of ionised carbon decreases as more carbon is locked up in
molecules, confirming that the two phases share the same location
\citep{sgg+91} and that the star formation rate may decrease with
increasing $ L_{\rm CO}$ \citep{sgg+91,pltv99,bji+00}.

\begin{figure}
 \centering
\includegraphics[angle=270,scale=0.75]{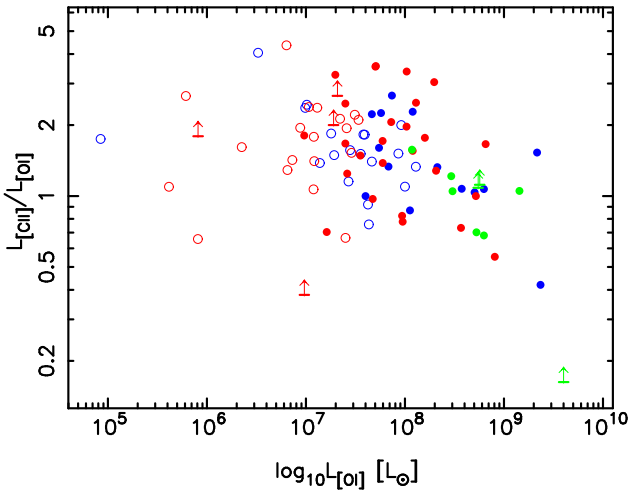}
\caption{As above, but normalising by the [O{\sc \,i}] luminosity.}
\label{cii-over-oi-oi}
    \end{figure}

The presence of molecular gas requires the presence of dust, the
heating of which will be responsible for the corresponding drop in
$L_{\rm [CII]}/L_{\rm FIR}$ with $L_{\rm FIR}$.  However, the changing
galactic demographics, due to the necessarily brighter sources at
larger luminosity distances, may have differing cooling mechanisms
than for the more proximate (and therefore dimmer) examples: The 63
$\mu$m [O{\sc \,i}] line has the effect of increasing the gradient of
the coolant line (\carb\ + [O{\sc \,i}]) luminosity (cf. $L_{\rm
  [CII]}$ only) against $L_{\rm FIR}$ and in Fig. \ref{cii-over-oi-oi}
we also see a decrease in $L_{\rm [CII]}/L_{\rm [OI]}$ with $L_{\rm
  [OI]}$.
In addition to this, \citet{piv07} suggest that the excited ($J =
4\rightarrow3$ \& $6\rightarrow5$. i.e. $\lambda = 651$ \& $434~\mu$m)
CO lines contribute as much to the cooling as the \carb\ line in one
of the ULIRGs (Mrk 231). From Fig.~\ref{cii-over-co-co} there is
undoubtably a steep increase in the relative CO luminosity, although
it is generally the $J = 1\rightarrow0$ transition which has been
detected in the ULIRGs and, again, these exhibit an excess in FIR
luminosity (see Fig.~\ref{cii-fir-co})\footnote{Also, \citet{piv07}
  believe that the lower rotational transitions do not trace the same
  gas phase.}. Note, however, that for the quasars, where the high
excitation transitions are redshifted into more ``observer friendly''
bands, the $J = 6\rightarrow5$ \& $7\rightarrow6$ ($\lambda = 434$ \&
$389~\mu$m) CO luminosities do follow the general trend
(Fig.~\ref{cii-fir-co}).

\begin{figure}
 \centering
\includegraphics[angle=270,scale=0.75]{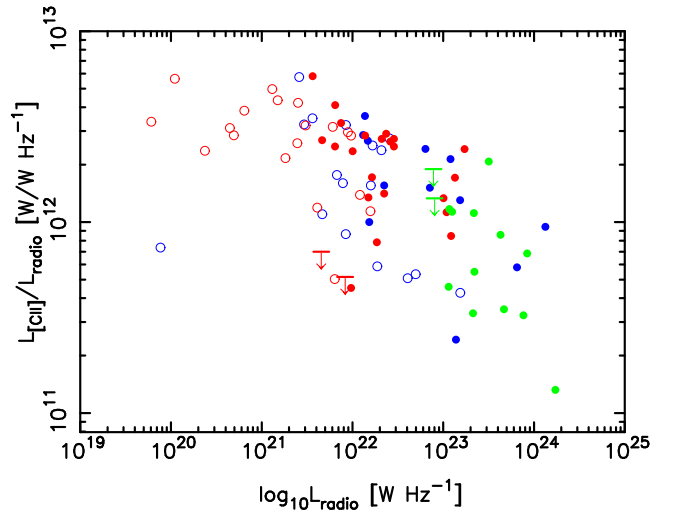}
\caption{As above, but normalising by the radio continuum luminosity.}
\label{cii-over-radio-radio}
    \end{figure}
The clear drop in $L_{\rm [CII]}/L_{\rm radio}$ with $L_{\rm radio}$
(Fig.~\ref{cii-over-radio-radio}), supports the possibility of a
changing AGN contribution. As discussed above, star forming galaxies
tend to be objects of low radio flux, whereas active galaxies give
rise to large radio fluxes and those of our sample, towards the high
end, certainly qualify as radio-loud. Quasars, or at least QSOs, are
often associated with substantial dust emission
(e.g. \citealt{sib97,bi02,cbk02}), as well as bright CO emission at
high redshift (see \citealt{hsy+04} and references therein). Again
this raises the possibility that the more radio luminous sources may
radiate the excess heat from the AGN through CO emission. Whatever
the cause, arguments involving the relative
decrease of $L_{\rm [CII]}$ with $L_{\rm FIR}$
\citep{mkh+01,nocr01,lsf+03}, must also account for similar decreases
as measured against the molecular gas and radio continuum
luminosities.

\section{Summary}

In addition to the well documented drop in the $L_{\rm [CII]}/L_{\rm
  FIR}$ ratio with far-infrared luminosity in extragalactic sources,
we find similar decreases with the molecular gas and radio continuum
luminosities. This indicates that there is a \carb\ deficit in
relation to each of these properties, which due to the flux limited
nature of the surveys, suggests a relative \carb\ decline with
luminosity. If evolutionary in nature, the order of
magnitude decrease in the mean $L_{\rm [CII]}/L_{\rm FIR}$ ratio over
the past 12\% of the history of the Universe, could be due to a decrease in
the metallicities, although, as per some of the literature, we see no
evidence of this. The decline in $L_{\rm [CII]}/L_{\rm FIR}$ is,
however, dominated by the ULIRGs at redshifts of $z\sim0.1$ as well as
the two QSOs, at $z=4.69$ and $6.42$, and so rather
than a detectable evolutionary effect, the decreasing $L_{\rm [CII]}/L_{\rm FIR}$
ratio is more likely the cause of a change in the demographics of the
objects which can be detected at these distances.

We suggest that the excess FIR and radio luminosities arise from
additional AGN activity, where the former is the result of dust in the
embedded circumnuclear torus being heated by ultra-violet photons, in
addition to the underlying ultra-violet emission from the stellar
population.  Both \citet{nocr01} and \citet{lsf+03} also advocate
non-PDR mechanisms as being responsible for some of the far-infrared
emission.  Furthermore, if the radio emission was due to the same
ionised gas as traced by the \carb\ emission, we would not expect a
relative decrease in $L_{\rm [CII]}$ with $L_{\rm radio}$. This may
also be indicative of an increasing AGN contribution to the luminosity
of these objects and towards the high end (i.e. the ULIRGs of
\citealt{lsf+03}), these sources would be considered radio galaxies.

Whether caused by an embedded AGN or vigorous star forming activity,
the heating of the dust is the most likely explanation for the
decrease in the \carb\ luminosity in relation to that of the FIR.
So although the gas is not expected to be heated by quite the same
extent as the dust, the changing galaxy types, as traced by the 
increasing luminosities, may imply different cooling mechanisms.
One possibility is an
increase in the relative contribution of the cooling by the [O{\sc
    \,i}] line, which is known to become more dominant at higher
ultra-violet fluxes \citep{kwhl99}, thus tracing the warmer dust
\citep{mkh+01}. Furthermore, like the molecular gas and radio
continuum luminosities, in comparison with the FIR, the [O{\sc \,i}]
also exhibits an excess over the \carb\ luminosity, although this is
also depleted for the ULIRGs. An additional coolant may therefore be
the CO emission, in which the higher rotational transitions are found
to rival the cooling by the \carb\ line in one of the ULIRGs of the
sample \citep{piv07}, with the two high redshift endpoints (for which
these transitions have also been observed) exhibiting no FIR excess in
relation to the CO.  We may therefore expect the warm molecular gas to
also be located in the torus from which the additional FIR luminosity
is arising, with the cooler gas, as traced by the low excitation
rotational transitions, located at larger radii. Beyond the
torus, in the main galactic disk, is also where
most of the cool neutral gas, as traced by the \HI\ 21-cm absorption,
is believed to reside (\citealt{cww08l} and references therein).

Observations of the higher rotational CO transitions in the ULRIGs
could determine whether these could contribute to the cooling budget
in these extreme FIR environments (due to starburst/AGN
activity), where the \carb\ emission is apparently lacking. As
well as this, our suggestion that the relative decline in \carb\ with
FIR is due to a changing demographic, and any associated evolutionary
effects, could be further tested by:

\begin{enumerate}
    \item Radio flux measurements of the two high redshift sources. At
      $z=4.69$ and $6.42$ the 1.4 GHz continuum flux would be
      redshifted to $250$ and $191$ MHz, respectively. Both of these
      frequencies are accessible by the Giant Metrewave Radio
      Telescope, although such low frequencies could be subject to
      severe interference.

    \item $[$O{\sc \,i}$]$ observations of these quasars in order
        to verify that the excess in this line over the \carb\ line
        extends beyond the local ($z\lapp0.13$) galaxies.

\item Confirming the $L_{\rm [CII]}/L_{\rm FIR}\lapp10^{-4}$ limit
    in PSS 2322+1944 at $z=4.12$, as referred to in \citet{mcc+05} [Benford et
      al., in prep.], but not since published. 

    \item Further observations of the \carb\ transition at $0.1296<z<4.69$,
      filling in the redshift gap in the $L_{\rm [CII]}/L_{\rm
        FIR}$--luminosity distance distribution (Fig. \ref{ratio-lbt}). At
      $z\gapp1$, sub-millimetre observations would also cover the redshift
      range where star formation is expected to be most prevalent
      \citep{pf95,llhc96}. However, at $\gapp400$~GHz these observations are
      difficult, requiring the very best atmospheric conditions.
      
\end{enumerate}

\section*{Acknowledgments}

I would like to thank the anonymous referee for their very helpful
comments which significantly improved the manuscript, as well as Matt
Whiting for commenting on an early draft. Also Matt
Whiting again, as well as Michael Murphy, for the various C subroutines I
utilise and Martin Thompson for debugging (my use of) these.  This
research has made use of the NASA/IPAC Extragalactic Database (NED)
which is operated by the Jet Propulsion Laboratory, California
Institute of Technology, under contract with the National Aeronautics
and Space Administration. This research has also made use of NASA's
Astrophysics Data System Bibliographic Service and {\sc asurv} Rev 1.2
\citep{lif92a}, which implements the methods presented in
\citet{ifn86}.


\end{document}